\pdfoutput=1

\documentclass[11pt]{article}

\usepackage[]{acl}

\usepackage{times}
\usepackage{latexsym}
\usepackage{graphicx}
\usepackage[T1]{fontenc}

\usepackage[utf8]{inputenc}

\usepackage{microtype}

\title{Better Transcription of UK Supreme Court Hearings }

\author{Hadeel Saadany\thanks{Email: \texttt{hadeel.saadany@surrey.ac.uk}} \\
Centre for Translation Studies \\
  University of Surrey \\
  United Kingdom \\\And
  Catherine Breslin\\
    Kingfisher Labs Ltd\\
    United Kingdom \\\AND
  Constantin Or\u{a}san \\
  Centre for Translation Studies\\
  University of Surrey \\
  United Kingdom \\\And
  Sophie Walker\\
  Just Access\\
  United Kingdom\\
  }

\begin{document}
\maketitle
\begin{abstract}

Transcription of legal proceedings is very important to enable access to justice. However, speech transcription is an expensive and slow process. 
In this paper we describe part of a combined research and industrial project for building an automated transcription tool designed specifically for the Justice sector in the UK.
We explain the challenges involved in transcribing court room hearings and the Natural Language Processing (NLP) techniques we employ to tackle these challenges. We will show that fine-tuning a generic off-the-shelf pre-trained Automatic Speech Recognition (ASR) system with an in-domain language model as well as infusing common phrases extracted with a collocation detection model can improve not only the Word Error Rate (WER) of the transcribed hearings but avoid critical errors that are specific of the legal jargon and terminology commonly used in British courts.


\end{abstract}

\section{Introduction}
\label{introduction}

There has been a recent interest in employing NLP techniques to aid in the textual processing of the legal domain \citep{elwany2019bert,nay_2021,mumcuouglu2021natural,frankenreiter2022natural}. In contrast, understanding spoken court hearings has not received the same attention as understanding the legal text documents. In the UK legal system, the court hearings sessions have a unique tradition of verbal argument. Moreover, these hearings crucially aid in new case preparation, provide guidance for court appeals, help in legal training and even guide future policy. However, the audio material for a case typically spans over several hours, which makes it both time and effort consuming for legal professionals to extract important information relevant to their needs. Currently, the existing need for legal transcriptions (covering 449K cases p.a in the UK across all court
tribunals \cite{Sturge:2021} is largely met by human transcribers.

\begin{table}[]
\centering
\begin{tabular}{lll}
\hline
\textbf{Model} &
  \textbf{Transcript} \\ \hline
\begin{tabular}[c]{@{}l@{}}Reference\\ AWS ASR\end{tabular} &
  \begin{tabular}[c]{@{}l@{}}So \textbf{my lady} um it is difficult to..\\ So \textbf{melody} um it is difficult to...\end{tabular} \\ \hline
\begin{tabular}[c]{@{}l@{}}Reference\\ AWS ASR\end{tabular} &
  \begin{tabular}[c]{@{}l@{}} it makes further \textbf{financial order}\\ it makes further \textbf{five natural}\end{tabular} \\ \hline
\end{tabular}

\caption{ Examples of Errors Produced by AWS ASR on Legal Hearings. Errors are typed in bold.}
\label{error1}
\end{table}

Although there are several current speech-to-text (STT) technology providers which could be used to transcribed this data automatically, most of these systems are trained on general domain data which may result in domain-specific transcription errors if applied to a specialised domain. 
One way to address this problem is for end-users to train their own ASR engines using their in-domain data.
However, in most of the cases the amount of data available is too low to enable them to train a system which can compete with well-knows cloud-based ASR systems which are trained on much larger datasets. 
In commercial scenarios, using generic cloud-based ASR systems to transcribe a specialised domain may result in a sub-optimal quality transcriptions for clients who require this service.

This holds particularly true for British court room audio procedures. When applying a generic cloud-based ASR system on British court rooms, the Word Error Rate (WER) remains relatively high due to long hearings, multiple speakers, complex speech patterns as well as unique pronunciation and vocabulary. Examples in Table \ref{error1} show common problems with transcribing speech from the legal domain using an on-the-shelf ASR systems such as AWS Transcribe\footnote{\url{https://aws.amazon.com/transcribe/}}. The references are taken from gold-standard edited transcripts of the UK Supreme Court Hearings\footnote{\url{https://www.supremecourt.uk/decided-cases/index.html}} created by the legal editors in our project. The first error is due to a special pronunciation of the phrase `\textit{my lady}' in British court rooms as it is pronounced like `\textit{mee-lady}' when barristers address a female judge. In the second example, the error is  related to legal terminology critical of the specific transcribed case.
Errors like in the second example are numerous in our dataset and affect also names and numbers. These errors can lead to serious information loss and cause confusion.

\begin{figure*}[!htbp]
\centering
  \scalebox{.6}{\includegraphics[scale=.5,trim={.2cm .2cm .3cm .2cm},clip]{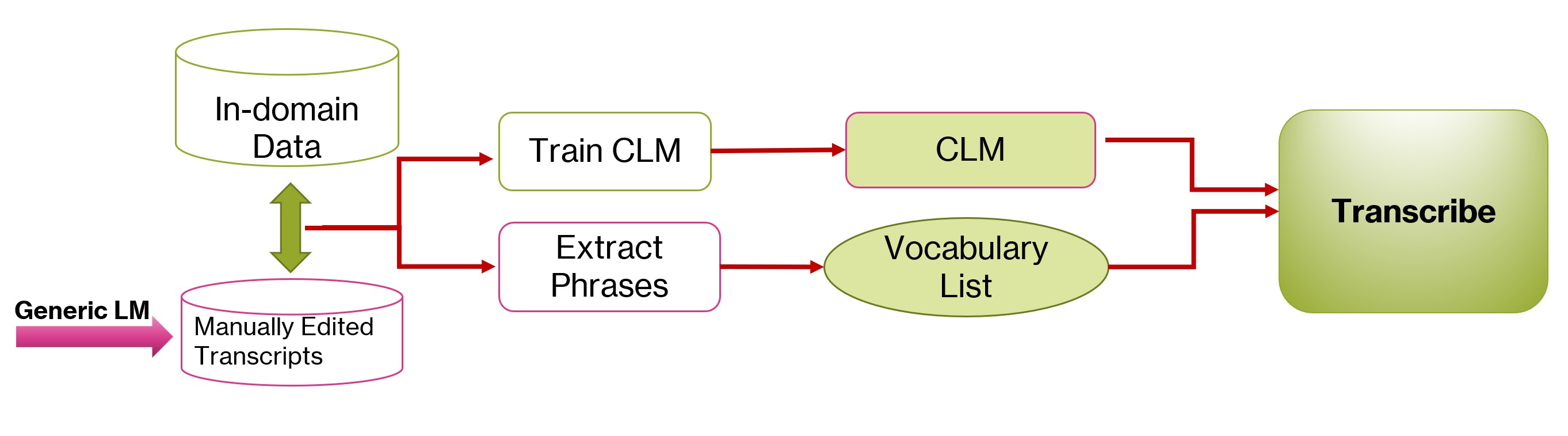}}
\caption{Pipeline for Improving ASR Output for Legal Specific Errors}
\label{fig:pipeline}
\end{figure*}

In this paper, we describe a joint research and commercial effort to perform domain adaptation of a generic ASR system to mitigate the errors in the automated UK court transcription services. We propose to minimise legal-specific errors by fine-tuning off-the-shelf ASR systems with a custom language model (CLM) trained on legal documents as well as 139 hours of gold-standard transcriptions of UK Supreme Court hearings. We also employ NLP techniques to automatically build a custom vocabulary of common multi-word expressions and word n-gram collocations that are critical in court hearings. We infuse our custom vocabulary to the CLM at transcription time. In this research, we evaluate the benefits of our proposed domain adaptation methods by comparing the WER of the CLM output with two off-the-shelf ASR systems: Amazon Web Services (AWS) Transcribe (commercial) and the OpenAI Whisper model (open-source) \citep{radfordrobust}. We also compare the general improvement in the ASR system's ability to correctly  transcribe legal entities with and without adopting our proposed methods.

\section{Related Work}
\label{sec:related work}
Automatic speech recognition (ASR) models convert audio input to text and they have optimal performance when used to transcribe data which is similar to the one they were trained on. However, performance degrades when there is a mismatch between the data used for training and the one that is being transcribed. Additionally, some types of audio material is intrinsically harder for speech recognition systems to transcribe. In practice, this means that speech recognition system performance degrades when, for example, there is background noise \citep{watanabe2020chime}, non-native accents \citep{feng2021quantifying,zhang2022mitigating}, young or elderly speakers \citep{feng2021quantifying}, or a shift in domain \citep{mai2022unsupervised}.


Performance degradation is typically mitigated by adapting or fine-tuning ASR models towards the domain of the targeted data by using a domain-specific dataset \citep{huo2021incremental,sato2022text,Dingliwa2022}. Some methods for domain adaptation adopt NLP techniques such as using machine translation models to learn a mapping from out-of-domain ASR errors to in-domain terms \citep{mani2020asr}. An alternative approach is to build a large ASR model with a substantially varied training set, so that the model is more robust to data shifts. An example of this latter approach is the recently released OpenAI Whisper model which is trained on 680k hours of diverse domain data to generalise well on a range of unseen datasets without the need for explicit adaptation \citep{radfordrobust}.


Moreover, ASR models are evaluated using Word Error Rate (WER), which treats each incorrect word equally. However, ASR models do not perform equally on different categories of words. Performance is worse for categories like names of people and organisations as compared to categories like numbers or dates \citep{del2021earnings}. ASR research targeted improving specific errors such as different named entities using  NLP techniques \citep{wang2020asr,das2022listen}. 

In this paper, we propose simple techniques to improve the effect of the domain mismatch between a generic ASR model and the specialised domain of British court room hearings. Our proposed method, improves both the system's WER rate as well as its ability to capture case-specific terms and entities. In the next section, we present our experiment set up and the evaluation results.  

\section{Experiment Set up}
Figure \ref{fig:pipeline} illustrates our proposed pipeline to improve ASR systems performance by legal domain-adaptation techniques. First, we build a custom language model (CLM) by fine-tuning a base ASR system, such as AWS Transcribe base system, using training data from the domain and a corpus of gold-standard legal transcriptions. Then, we use NLP techniques to extract domain-specific phrases and legal entities from the in-domain data to create a vocabulary list. We use both the CLM and the vocabulary list for transcribing legal proceedings. The following sections explain details of our experiment where we implemented this pipeline on the AWS Transcribe base model. We compare the performance of our CLM model with different settings to AWS Transcribe base ASR system and OpenAI Whisper open-source ASR system when transcribing $\approx$ 12 hours of UK Supreme Court Hearings.


\subsection{Dataset}

For training our CLM, we use two datasets from the legal domain. The first is  Supreme Court written judgements of 43 cases consisting of  3.26M tokens scraped from the official site of the UK Supreme Court\footnote{https://www.supremecourt.uk/decided-cases/}. The second dataset consists of $\approx$ 81 hours of gold-standard transcripts of 10 Supreme Court hearings. The gold-standard transcripts are created by  post-editing the AWS Transcribe output of the court hearings by a team of legal professionals using a specially designed interface. We use both datasets to train CLM on the AWS platform.

For the vocabulary list, we use a dataset of $\approx$ 139 hours of gold-standard transcriptions of Supreme Court hearings along with the supreme court judgements used for training the CLM. To extract the vocabulary from this dataset, we implement two methods. First, we use this dataset to train a phrase detection model that collocates bigrams based on Pointwise Mutual Information (PMI) scoring of the words in context \citep{mikolov2013distributed}\footnote{We train the model using the Gensim Python library \citep{rehurek_lrec}}. Then, the collocation model is used to extract a list of most common bigrams in the dataset. This list includes frequent legal terms and phrases specific of the Supreme Court cases included in the training corpus. Second, we use Blackstone\footnote{https://research.iclr.co.uk/blackstone}, an NLP library for processing long-form and unstructured legal text, to  extract a list of legal entities from the dataset. The list of  legal entities included: Case Name, Court Name, Provision (i.e. a clause in a legal instrument), Instrument (i.e. a legal term of art) and Judge. We concatenated this Blackstone entity list with the spaCy v3.4 library list of non-legal entities such as: Cardinals, Persons and Dates. The results of applying our methods for the transcription of 2 Supreme Court case hearings consisting of 12 hours is explained in the next section.

\section{Results}
\label{results}

\begin{table*}[]
\centering

\begin{tabular}{|l|l|l|l|l|}
\hline
\textbf{Model} &
  \textbf{\begin{tabular}[c]{@{}l@{}}WER\\ Case1\end{tabular}} &
  \textbf{\begin{tabular}[c]{@{}l@{}}WER\\ Case2\end{tabular}} &
  \textbf{\begin{tabular}[c]{@{}l@{}}WER\\ Average\end{tabular}} &
  \textbf{Transcription Time} \\ \hline
\textbf{AWS base}    & 8.7          & 16.2          & 12.3          & 85 mins \\ \hline
\textbf{CLM1}       & 8.5          & 16.5          & 12.4          & 77 mins \\ \hline
\textbf{CLM2}        & \textbf{7.9} & 15.5          & \textbf{11.6} & 77 mins \\ \hline
\textbf{CLM2+Vocab}  & \textbf{7.9} & 15.6          & \textbf{11.6} & 132 mins \\ \hline
\textbf{CLM2+Vocab2} & 8.0             & 15.6          & 11.7          & 112 mins \\ \hline
\textbf{Whisper}              & 9.6          & \textbf{15.3} & 12.4          & 191 mins  \\ \hline
\end{tabular}
\caption{Average WER and Transcription Time}
\label{tab:WER}
\end{table*}




\begin{table*}[!htbp]
\centering

\begin{tabular}{|l|c|c|c|}
\hline
\textbf{Entity} & \textbf{AWS BASE} & \textbf{Whisper}&  \textbf{CLM2+vocab}  \\ \hline
\textbf{Judge}  & 0.66 & 0.77 & 0.84  \\ \hline
\textbf{CASE NAME}  & 0.69 & 0.85 &  0.71 \\ \hline
\textbf{Court}      & 0.98 &1  & 0.93 \\ \hline
\textbf{Provision}    & 0.88 & 0.95 & 0.97\\ \hline
\textbf{Cardinal} & 1 & 0.97  &  1 \\ \hline
\end{tabular}
\caption{Ratio of Correctly Captured Legal Entities by the ASR Systems}
\label{tab:ratio}
\end{table*}

Table \ref{tab:WER} shows the WER scores and WER average score for the 2 transcribed cases with different CLM system settings, as well as, for the two baseline systems: the AWS Transcribe (AWS base) and Whisper. The different CLM settings are as follows: CLM1 is trained on only the texts of the Supreme Court judgements, CLM2 is trained on both the judgements and the gold-standard transcripts, CLM2+Vocab uses CLM2 for transcription plus the global vocabulary list extracted by our phrase detection model, and CLM2+Vocab2 uses CLM2 for transcription plus the legal entities vocabulary list extracted by Blackstone library. 

As can be seen in Table \ref{tab:WER}, the ASR performance is consistently better with the CLM models than with the generic ASR systems for the two transcribed cases. CLM2 model, trained on textual data (i.e. the written judgements) and gold-standard court hearing transcriptions, outperforms AWS base and Whisper with a 9\% and 8\% WER improvement, respectively. Moreover, we observe around 9\% improvement in average WER score over the two generic models when concatenating  the list of legal phrases that is extracted by our phrase detection model with the CLM2 system. While ASR error correction  indicates an improved transcription quality with our proposed domain adaptation methods, we also evaluated the ASR systems performance with specific errors such as legal entities and terms.

Table \ref{tab:ratio} shows the average ratio of correctly transcribed legal entities in the two studied court room hearings. We compare the performance of CLM2 infused with the legal terms list (CLM2+Vocab) to the two generic ASR systems. The ratios in Table \ref{tab:ratio} indicate that CLM2+Vocab is generally more capable of transcribing legal-specific terms than the other two models. It is also better at transcribing critical legal entities such as Provisions.\footnote{A Provision, a statement within an agreement or a law, typically consists of alphanumeric utterances in British court hearings (e.g. `section 25(2)(a)-(h)' or `rule 3.17').} Such legal terminology needs to be accurately transcribed. Our CLM2 model with legal vocabulary demonstrates better reliability in transcribing these terms. 

A similar trend is evident with the legal entity Judge which refers to the forms of address used in British court rooms (e.g. `Lord Phillips', `Lady Hale'). This entity is typically repeated in court hearings whenever a barrister or solicitor addresses the court.  We see that both the generic ASR systems perform badly on this category with ratios of 0.66 and 0.69, respectively. On the other hand, we observe a significant improvement in correctly transcribing this entity by the CLM2+Vocab with a ration of 0.84 correct transcriptions.

In addition to evaluating the output of the ASR engines, we also recorded the time required to produce the transcription. The models based on AWS were run in the cloud using the Amazon infrastructure. Whisper was run on a Linux desktop with an NVIDIA GeForce RTX 2070 GPU with 8G VRAM. For all the experiments, the medium English-only model was used. As expected the fastest running time is obtained using the AWS base model. Running the best performing model increases the time by 155\%, whilst Whisper more than doubles it.

\section{Conclusion}

In this paper, we present a study to show the effect of domain adaption methods on improving the off-the-shelf ASR system performance in transcribing a specialised domain such as British court hearings. We optimised the performance of the ASR system by training an ASR custom language model on gold-standard legal transcripts and textual data from the legal domain. We also trained a phrase detection model to incorporate extracted list of data-specific bigram collocations at transcription time. We evaluated the ASR quality improvements both in terms of average WER and ratio of correctly transcribed legal-specific terms. We observe significant gains in the ASR transcription quality by our domain adaptation techniques. For commercial use of ASR technologies, improving error rate in general and transcription quality of critical legal terms in particular would minimise manual post-editing effort and hence save both time and money. We plan to evaluate the impact of different configurations proposed in this paper on the editors' postediting effort. 

In the future, we will expand to record data from a variety of accents to address another axis of degradation in British audio procedures different than the Supreme Court hearings which are mostly a homogeneous group of speakers. We will also explore the ability to use NLP topic modelling techniques to connect legal entities that were crucial in a court's case decision.

\bibliography{main}

\end{document}